# Precision Agriculture: Ultra-Compact Sensor and Reconfigurable Antenna for Joint Sensing and Communication

Ali Raza[1, 2, 3], Rasool Keshavarz[1], Negin Shariati[1, 2]

*Abstract*— In this paper, a joint sensing and communication system is presented for smart agriculture. The system integrates an Ultra-compact Soil Moisture Sensor (UCSMS) for precise sensing, along with a Pattern Reconfigurable Antenna (PRA) for efficient transmission of information to the base station. A multi-turn complementary spiral resonator (MCSR) is etched onto the ground plane of a microstrip transmission line to achieve miniaturization. The UCSMS operates at 180 MHz with a 3-turn complementary spiral resonator (3-CSR), at 102 MHz with a 4-turn complementary spiral resonator (4-CSR), and at 86 MHz with a 5-turn complementary spiral resonator (5-CSR). Due to its low resonance frequency, the proposed UCSMS is insensitive to variations in the Volume Under Test (VUT) of soil. A probe-fed circular patch antenna is designed in the Wireless Local Area Network (WLAN) band (2.45 GHz) with a maximum measured gain of 5.63 dBi. Additionally, four varactor diodes are integrated across the slots on the bottom side of the substrate to achieve pattern reconfiguration. Six different radiation patterns have been achieved by using different bias conditions of the diodes. In standby mode, PRA can serve as a means for Wireless Power Transfer (WPT) or Energy Harvesting (EH) to store power in a battery. This stored power can then be utilized to bias the varactor diodes. The combination of UCSMS and PRA enables the realization of a joint sensing and communication system. The proposed system's planar and simple geometry, along with its high sensitivity of 2.05 %, makes it suitable for smart agriculture applications. Moreover, the sensor is adaptive and capable of measuring the permittivity of various Material Under Test (MUT) within the range of 1 to 23.

## I. INTRODUCTION

ACCURATE measurement of soil moisture levels in agriculture is crucial for enhancing the quality and quantity of crop yields. In the realm of smart agriculture, precise monitoring and effective management of soil moisture play a pivotal role in optimizing irrigation practices, addressing water scarcity challenges, and maximizing agriculture productivity [1, 2]. Conventional methods of soil moisture measurement are labor-intensive and time-consuming. Therefore, the deployment of soil moisture sensors in a wireless sensor network (WSN) as a part of Internet-of-Things (IoT) has gained significant attention in smart agriculture [3].

Microwave sensors have been widely presented for soil moisture measurements in recent years [4-10]. Various techniques have been utilized, including frequency domain reflectometry (FDR) [4, 5, 11], time domain reflectometry (TDR), time domain transmission (TDT) [6, 12, 13], and capacitive sensing [7-10], to measure the permittivity of the material under test (MUT). In capacitive sensors, the value of capacitance is measured by using the discharge time of the capacitor through a resistor, which is then used to determine the permittivity of the MUT. However, variations in temperature can affect the resistance, leading to inaccurate measurements. The permittivity of the soil can be correlated to its moisture content [14]. The volumetric water content (VWC) quantifies the amount of water present in the soil and is calculated using (1), where $W_1$ and $W_2$ represent the weights of the dry soil and water, respectively.

$$VWC\ (\%) = \frac{W_2}{W_1 + W_2} \times 100 \qquad (1)$$

Some previously reported sensors solely relied on the real value of permittivity ($\varepsilon'_m$) to calculate the VWC, disregarding the contribution of the imaginary part ($\varepsilon''_m$) [4]. However, the $\varepsilon''_m$ of the soil varies from 0.05 to 3.5 at 130 MHz, corresponding to a VWC range of 0–30 %. Neglecting the contribution of $\varepsilon''_m$ can lead to considerable errors in soil moisture measurement.

The theory behind low-frequency RF signals suggests that they can penetrate deeper into the soil compared to high-frequency signals [15]. Consequently, a low-frequency FDR-based sensor can provide measurements of soil moisture at greater depths, making it a more suitable choice for practical environments where moisture distribution is uneven across the soil. In high-temperature conditions, surface soil moisture tends to be lower, while deeper soil layers retain higher moisture levels. Therefore, a high-frequency sensor may produce inaccurate measurements due to its limited signal penetration depth. However, using lower resonance frequencies for the signal necessitates larger sensor structures, which can pose handling challenges.

[1]A. Raza, R. Keshavarz and N. Shariati are with RF and Communication Technologies (RFCT) research laboratory, School of Electrical and Data Engineering, Faculty of Engineering and IT, University of Technology Sydney, Ultimo, NSW 2007, Australia. (e-mail: Ali.Raza-1@student.uts.edu.au)

[2]A. Raza and N. Shariati are also with Food Agility CRC Ltd, Sydney, NSW, Australia 2000.
[3]A. Raza is also with the Department of Electrical, Electronics and Telecommunication Engineering, University of Engineering and Technology (UET), Lahore, Punjab, Pakistan.

Metamaterial Transmission Lines (MTLs) have been widely used to minimize the size of microwave devices, such as antennas [16-18], power dividers [19], filters [20], and resonators [21]. Within MTLs, split-ring resonators (SRRs) and complementary split-ring resonators (CSRRs) have been extensively adopted to achieve size reduction [22]. Various SRR and CSRR-based microwave sensors are reported in the literature [4, 23-28]. An FDR based differential sensor for soil moisture measurements is presented in [4]. The sensor utilizes both SRR and CSRR on opposite sides of the substrate to measure the permittivity of the MUT. However, the sensor exhibits low sensitivity, and its resonance frequency is high (4 GHz), limiting its coverage area for soil moisture measurements. Another differential sensor based on SRR is presented in [11]. The sensor operates at a high frequency (5.12 GHz), which makes it susceptible to variations in the Volume Under Test (VUT). Due to these frequency variations with different VUTs, the reported sensor can generate false measurements. Low frequency FDR sensor are presented in [5, 29]. The sensors operate at 560 MHz and 1.017 GHz when unloaded; however, the structures are large, and the sensors have low sensitivity. Several microwave sensors have been reported for the characterization of microfluids [24, 25, 27, 30]. These sensors operate at frequencies of 2.4 GHz, 2.45 GHz, 2.234 GHz, and 2.38 GHz, respectively. While these sensors have the ability to measure high permittivity values as high as 70, they also exhibit very low sensitivities. Several high-sensitivity CSRR-based sensors have also been proposed for measuring the permittivity of the MUT [23, 26, 28]. However, significant frequency variations occur for different heights of the MUT, primarily because of the high resonance frequencies of these sensors. Recently, a permittivity sensor and an antenna are integrated to enable sensing and communication [31]. The antenna is designed to operate at 2.45 GHz Wireless Local Area Network (WLAN) band, while the sensors' resonance frequency is 4.7 GHz. A frequency-selective multipath filter is utilized to measure the permittivity of the MUT to characterize the material. However, the sensor has a high resonance frequency and does not produce narrow resonances, which can lead to false measurements. Designing a sensor with low resonance frequency, compact size, and high sensitivity to cover a large VUT is highly challenging.

For remote sensing systems in diverse geographical structures, it is essential to integrate an antenna for transmitting sensor data to the base station. This enables remote monitoring of real-time soil moisture data for precise irrigation and enhances resource management efficiency. Agricultural areas would have different geographical structures and the positioning of IoT equipment can vary. A large-coverage antenna becomes particularly essential in agricultural areas where geographical structures vary, and the positioning of IoT equipment may differ. An omnidirectional antenna has the ability to communicate in all directions and can be used for communication with the base station regardless of the positioning. However, the gain of an omnidirectional antenna is very low, which makes it unsuitable for large agricultural fields. Therefore, a directional antenna with pattern reconfiguration is required for reliable long-distance communication in various geographical structures. Various pattern reconfigurable antennas are reported in the literature to switch the radiation pattern of the antenna using radio-frequency microelectromechanical system (RF MEMS) switches [32], PIN diodes [33-42], and varactor diodes [43]. A slot-based electronically steerable pattern reconfigurable array is presented for IoT applications in [44]. It utilizes six pin diodes to achieve pattern reconfiguration of a monopole antenna in the WLAN band, but the geometry is complex and non-planar. Another pattern reconfigurable antenna is presented in the WLAN band for IoT applications [45]. The antenna consists of 4 wire patch antennas, and the reconfiguration is achieved by using a single single-pole-four-throw (SP4T). However, the design is non-planar and requires a microcontroller to control the SP4T switch. A probe-fed varactor-loaded pattern reconfigurable antenna is presented for continuous beam steering, utilizing four varactor diodes [46]. However, the antenna's geometry is multi-layer.

Designing a sensor with a low resonance frequency to cover a large VUT while maintaining a compact size and high sensitivity is highly challenging. In the literature, various types of FDR-based sensors have been reported for permittivity measurement. Some sensors operate at low frequencies but have large sizes, while others exhibit high sensitivities but with limitations in measuring high permittivity values. For practical remote sensing systems in diverse geographical structures, it is also essential to integrate a pattern reconfigurable antenna for transmitting sensor data to the base station. Thus, the objective of this research was to develop a compact sensor with both low resonance frequency and high sensitivity for precise sensing, along with a pattern reconfigurable antenna for effective data transmission in various geographical locations.

In this paper, we propose a sensing and communication system for smart agriculture. The proposed system comprises an Ultra-compact Soil Moisture Sensor (UCSMS) and a radiation Pattern Reconfigurable Antenna (PRA) for both sensing and communication purposes. The proposed UCSMS operates at a low resonance frequency to cover a larger VUT and provide in-depth soil moisture sensing. The sensor is designed using a multi-turn complementary spiral resonator (MCSR) in the ground plane of a microstrip transmission line to realize a miniaturized and planar structure. The UCSMS operates at a low frequency of 86 MHz with a 5-turn complementary spiral resonator (5-CSR). This results in a sensor with a compact size ($0.028 \times 0.028 \ \lambda_0{}^2$), low resonance frequency, and high sensitivity, making it well-suited for soil moisture measurements. A directional antenna with pattern reconfiguration is designed at 2.45 GHz WLAN band for reliable long-distance joint communication for diverse geographical structures. The proposed PRA is designed using CSRR and U-shaped slots and to achieve pattern reconfiguration, four varactor diodes are integrated with the CSRRs. By changing the biasing voltage of the diodes, six different radiation patterns can be achieved. Moreover, this design also enables the utilization of the proposed PRA for wireless power transfer (WPT) or energy harvesting (EH) in the 2.45 GHz WLAN band to store power in a battery. This stored power can then be utilized to power the varactor diodes. The precise sensing capability and multi-direction communication feature of the proposed system make it suitable for precision agriculture applications.

The major contributions of this paper are summarized as follows:

1. The proposed UCSMS operates at a low 170 MHz frequency with a 3-CSR, offering high sensitivity. This makes it suitable for in-depth sensing and for covering a large volume of soil, ensuring accurate soil moisture measurements. Covering a larger volume reduces the need for numerous sensors on a large-scale farm, resulting in cost savings for farmers and simpler implementation.

2. Based on the UCSMS's performance in an Environmental Testing Chamber (Temperature and Humidity), it is well-suited for accurate measurements in real-world agricultural settings with variable temperature and humidity.

3. In this article, the UCSMS is primarily used for soil moisture sensing. However, the proposed UCSMS is adaptive and has the ability to measure material permittivity within the range of 1–23.

4. The proposed PRA has six different radiation modes, enabling the generation of six radiation patterns with different biasing conditions. The proposed PRA operates at 2.45 GHz Wireless Local Area Network (WLAN) band and offers a broad coverage area along with high gain. Consequently, the same antenna can serve as a receiving antenna for EH or WPT to store power in a battery.

5. The integration of UCSMS and RPA makes the proposed system suitable for smart agriculture. By sensing complex permittivity to achieve precise VWC measurements, it can then transmit this data to the base station, thus enabling remote monitoring of soil moisture.

The organization of this paper is as follows: Section II covers the working principle, design, and theoretical analysis of the UCSMS, as well as the PRA design. Section III presents the simulated and measured results of the UCSMS and PRA. Finally, Section IV presents the conclusion.

## II. DESIGN METHODOLOGY

The working principle, design, and theory of the proposed system are discussed in the following subsections:

### A. Working Principle of the System

The proposed joint sensing and communication system comprises a UCSMS and a PRA designed for smart agriculture. Fig. 1 illustrates the working principle of the proposed system.

A practical setup for sensing and communication is shown in Fig. 1(a), where a continuous wave RF signal is transmitted to the UCSMS through a circulator, and the reflected signal is then measured at port-3 of the circulator using a power detector. The frequency of the reflected signal is used to measure the complex permittivity of the MUT, which corresponds to a specific VWC. This information is transmitted to the base station using a directional pattern reconfigurable antenna. The PRA is designed with the capability to change the directions of the radiation pattern for various geographical structures. In standby mode, the PRA can be utilized for WPT or EH. In the experimental setup shown in Fig. 1(b), a vector network analyzer (VNA) is utilized to measure the frequency responses for different VWC levels using UCSMS. This proposed UCSMS can be utilized to measure the permittivity of various materials by configuring the system as depicted in Fig. 1.

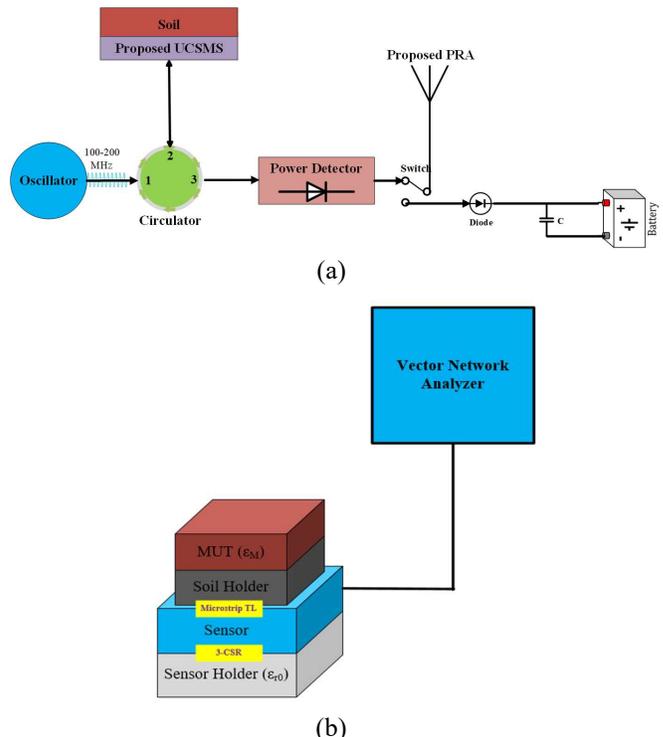

Fig. 1. Working principle of the proposed system, (a) working principle of the combined system, (b) laboratory setup for testing of the UCSMS.

### B. UCSMS Design

The proposed ultra-compact soil moisture sensor (UCSMS) is designed using an FR-4 substrate ($\varepsilon_r = 4.3$, $tan\delta = 0.025$, $h = 1.6$ mm) and has dimensions of $50 \times 50$ mm² as shown in Fig. 2. The proposed sensor incorporates an MCSR and a microstrip feed line for exciting the resonator. An open circuit stub of 5 mm length is connected at a distance of 16 mm with the transmission line to achieve a better impedance matching for a wide range of permittivity ($\varepsilon'_r$) values. The reflection coefficients of the proposed structure with and without the open circuit stub are shown in Fig. 3.

The resonance frequency of the spiral resonator can be calculated using equivalent inductance and capacitance (2). Increasing the number of turns in the resonator results in a corresponding increase in the equivalent inductance while reducing the width of the turns increases the equivalent capacitance [22]. To analyze different configurations of the spiral resonator, simulations were conducted by varying the number of turns and the width of the turns. Fig. 4 illustrates three different configurations of the spiral resonator, along with their corresponding equivalent circuits. It can be observed from the equivalent circuits that increasing the number of turns and/or decreasing the width of the turns will increase the equivalent inductance and/or capacitance, resulting in the reduction of the resonance frequency. However, the practical limitation of the fabrication equipment makes it challenging to reduce the width and spacing beyond a certain limit. A comparison between the 3-turn complementary spiral resonator (3-CSR) with a turn width ($w_t$) of 1 mm, the 4-turn complementary spiral resonator (4-CSR) with a $w_t$ of 0.5 mm, and 5-turn complementary spiral resonator (5-CSR) with a $w_t$ of 0.5 mm is presented in Fig. 5. The simulated resonance frequencies of 3-CSR, 4-CSR, and 5-CSR are 180 MHz, 102 MHz and 86 MHz, respectively, which validates the aforementioned theory. For this study and to prove the concept, the 3-CSR was selected, fabricated, and tested as a soil moisture sensor.

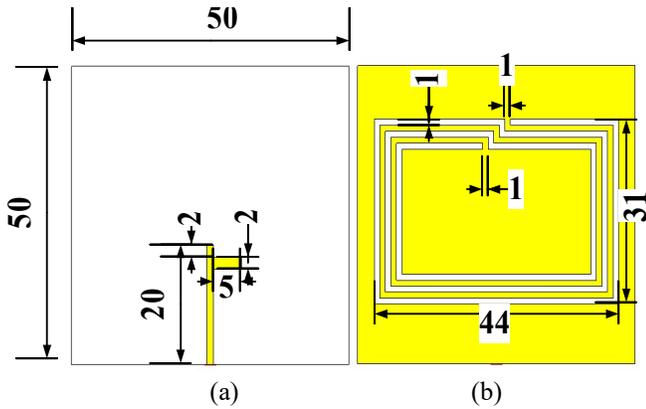

Fig. 2. Geometry of the proposed UCSMS, (a) top view, (b) bottom view, (all dimensions are in mm).

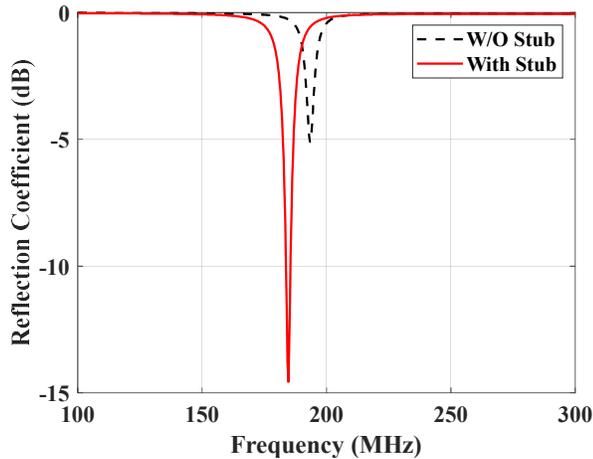

Fig. 3. Reflection coefficient of the proposed UCSMS without and with open stub.

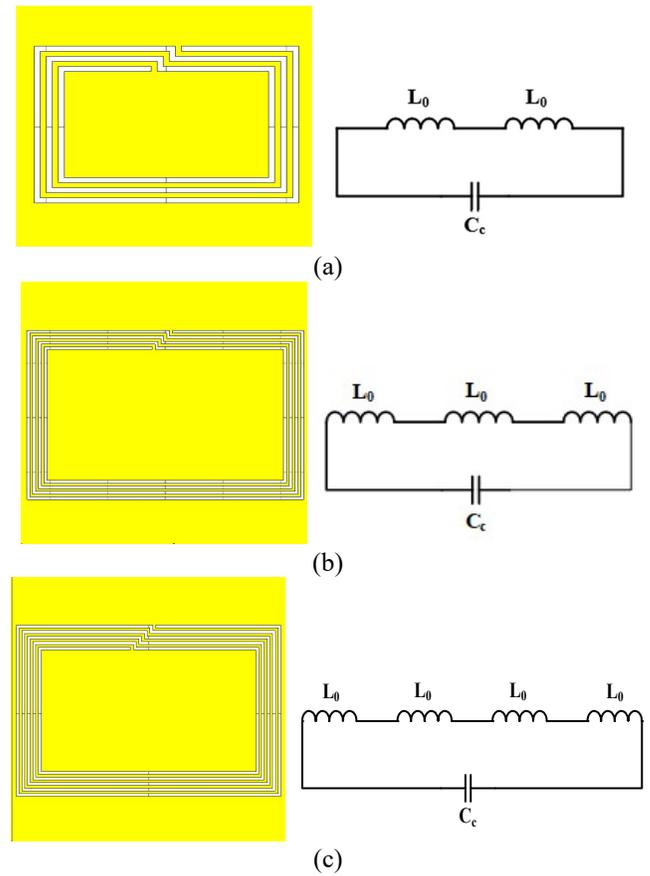

Fig. 4. Geometry and equivalent circuits of different complementary spiral resonators, (a) 3-CSR, (b) 4-CSR, (c) 5-CSR.

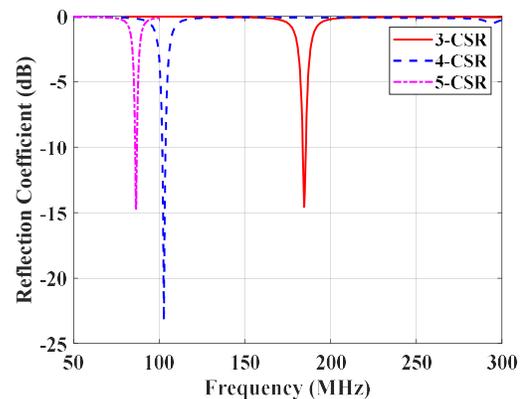

Fig. 5. Comparison of different complementary spiral resonators.

$$f = \frac{1}{2\pi\sqrt{L_e C_e}} \qquad (2)$$

The equivalent circuit of a spiral resonator is an *LC* tank circuit as illustrated in Fig. 4. For instance, the equivalent circuit of 5-CSR consists of four inductances and a capacitance connected in series. The inductance value of the resonator can be calculated using (3) [47], where $Z_0$ is the line impedance, $\varepsilon_e$ is the effective permittivity, $c$ is the speed of light, and $l$ is the line length.

$$L_0 = \frac{Z_0 \sqrt{\varepsilon_e}}{c} l \qquad (3)$$

The line impedance ($Z_0$) and the effective permittivity ($\varepsilon_e$) can be calculated using (4) and (5) [48], where $h$ represents the thickness of the substrate and $w_t$ is the width of a single turn. The thickness of the substrate is 1.6 mm.

$w_t < h$:

$$\varepsilon_e = \frac{\varepsilon_r + 1}{2} + \frac{\varepsilon_r + 1}{2}\left[\frac{1}{\sqrt{1 + 12\left(\frac{h}{W_t}\right)}} + 0.04\left(1 - \frac{W_t}{h}\right)^2\right] \quad (4.a)$$

$$Z_0 = \frac{60}{\varepsilon_e}log_2\left[8\frac{h}{W_t} + 0.25\frac{W_t}{h}\right] \quad (4.b)$$

$w_t > h$:

$$\varepsilon_e = \frac{\varepsilon_r + 1}{2} + \frac{\varepsilon_r + 1}{2}\left[\frac{1}{\sqrt{1 + 12\left(\frac{h}{W_t}\right)}}\right] \quad (5.a)$$

$$Z_0 = \frac{120\pi}{\varepsilon_e\left[\frac{W_t}{h} + 1.393 + \frac{2}{3}log_2\left(\frac{W_t}{h} + 1.444\right)\right]} \quad (5.b)$$

### C. PRA Design

After measuring the soil moisture using UCSMS, an antenna is required to transmit this information to the base station to realize smart agriculture. A directional pattern reconfiguration antenna is designed for reliable long-distance communication in various geographical structures.

The proposed PRA is designed on a Rogers RO4003C substrate ($\varepsilon_r = 3.55$, $tan\delta = 0.0027$). The PRA geometry consists of a coaxial probe-fed circular patch, two CSRRs, and two U-shaped slots. Each CSRR slot has a commentary split-ring resonator and a coplanar waveguide (CPW) stub. The circular patch is positioned on the top side of the substrate, while the CSRRs and U-shaped slots are located on the bottom side. The structure is symmetrical along the *x-axis* and *y-axis* and the size of the substrate is $L_s \times W_s \times H$ mm$^3$. Both left and right CPW-based CSRRs are responsible for the radiation pattern in the *xz-plane*, while top and bottom U-shaped slots are employed for the radiation pattern in the *yz-plane*. To achieve pattern reconfiguration in the *xz-plane* while maintaining the same *yz* pattern, two varactor diodes are integrated across each symmetrical CPW-based CSRR. The PRA is excited using a coaxial probe at a distance of L4 to achieve better impedance matching.

The detailed geometry of the proposed PRA is shown in Fig. 6, with Fig. 6(a) depicting the top view, Fig. 6(b) showing the bottom view, and Fig. 6(c) providing a magnified view. Four varactor diodes are integrated across both symmetrical CPW-based CSRRs at an optimized position to achieve pattern reconfiguration in the *xz-plane*. A biasing network is designed and connected to the varactor diodes using vias, as shown in Fig. 6(a). The antenna is designed for the 2.45 GHz WLAN band, and the optimized dimensions of the structure are provided in Table 1. By selecting suitable biasing voltages, six different radiation patterns have been generated in various directions, named as 'front', 'back', 'upper left', 'left', 'upper right', and 'right', as shown in Fig. 7. TABLE 2 summarizes the biasing voltages and corresponding capacitance values for the six different radiation patterns of the proposed PRA. In the simulation, the varactor diode is modeled as a non-linear circuit and a parametric sweep is utilized to change the value of the capacitance.

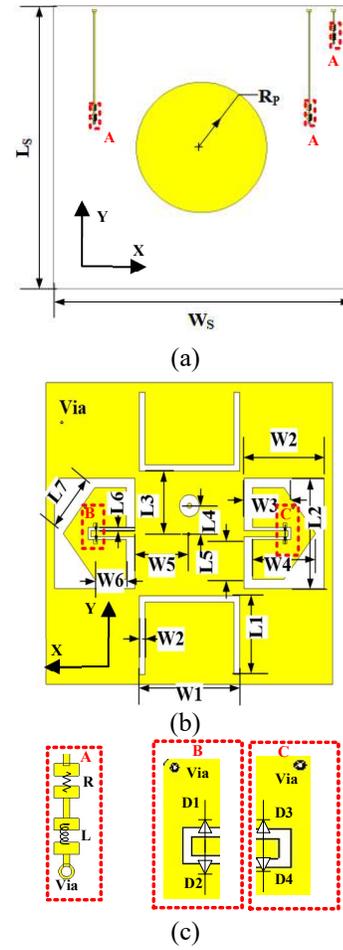

Fig. 6. Geometry of the proposed antenna, (a) top view, (b) bottom view, (c) magnified view of areas A, B, and C.

TABLE 1. DIMENSIONS OF THE PROPOSED PATTERN RECONFIGURABLE ANTENNA (ALL UNITS ARE IN MM)

| $L_S$ | $W_S$ | $R_P$ | L1 | W1 | W2 | L2 |
|---|---|---|---|---|---|---|
| 81.175 | 85 | 18.7 | 21.25 | 29.75 | 1.7 | 29.75 |
| **L3** | **L4** | **W2** | **W3** | **W4** | **L5** | **W5** |
| 16.8 | 7.53 | 23.8 | 13 | 18.7 | 10.625 | 16.15 |
| **L6** | **W6** | **L7** | **H** | | | |
| 0.85 | 9.35 | 15.47 | 1.524 | | | |

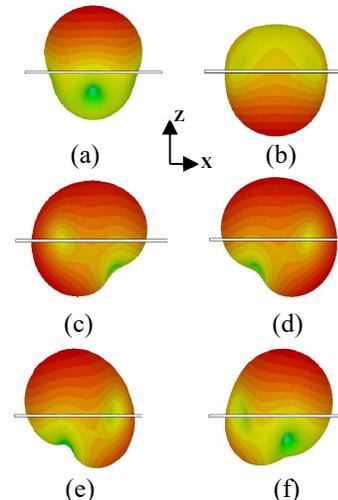

Fig. 7. Radiation patterns of the proposed PRA at 2.45 GHz: (a) front, (b) back, (c) left, (d) right, (e) upper left, (f) upper right.

TABLE 2. BIASING VOLTAGES AND CAPACITANCE VALUES OF VARACTOR DIODES FOR DIFFERENT RADIATION PATTERNS

| Varactor Diodes (D1, D2) | | Varactor Diodes (D3, D4) | | Radiation Pattern |
|---|---|---|---|---|
| Bias Voltage (V) | Capacitance (pF) | Bias Voltage (V) | Capacitance (pF) | - |
| 0 | 2.35 | 0 | 2.35 | Front |
| 0 | 2.35 | 3 | 0.970 | Upper left |
| 0 | 2.35 | 15 | 0.466 | Left |
| 3 | 0.970 | 0 | 2.35 | Upper right |
| 15 | 0.466 | 0 | 2.35 | Right |
| 15 | 0.466 | 15 | 0.466 | Back |

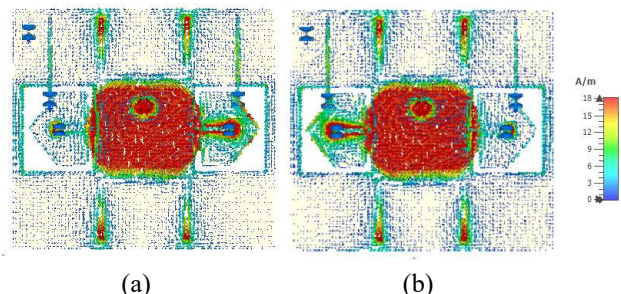

(a)                    (b)

Fig. 8. Current distribution of the PRA, (a) left, (b) right.

The surface current distributions for the 'left' and 'right' directions are shown in Fig. 8. For the 'left' case, the diodes D1 and D2 have a capacitance value of 2.35 pF, resulting in low reactance at 2.45 GHz. Due to this low reactance value, the current bypasses the right CSRR and instead travels through the diodes as shown in Fig. 8(a). However, the diodes D3 and D4 have a capacitance value of 0.466 pF, resulting in a higher reactance value for the current at 2.45 GHz. As a result, the current flows through the left CSRR and produces the 'left' radiation pattern. For the 'right' case, diodes D1 and D2 have a capacitance value of 0.466 pF, while diodes D3 and D4 have a capacitance value of 2.35 pF. This configuration results in the generation of the 'right' radiation pattern.

## III. SIMULATION AND MEASUREMENT RESULTS

### A. Simulated and Measured Results of UCSMS:

To showcase the performance of the proposed UCSMS, the MCSR sensor is designed and simulated using CST MWS 2019. Furthermore, the optimized UCSMS with 3-CSR is fabricated on an FR-4 substrate, and the prototype is shown in Fig. 9. The simulated and measured reflection coefficients of the unloaded UCSMS with 3-CSR are shown in Fig. 10. The measured resonance frequency of the sensor is 170 MHz with a narrow bandwidth, and a close agreement between the simulated and measured results indicates the validation of the structure.

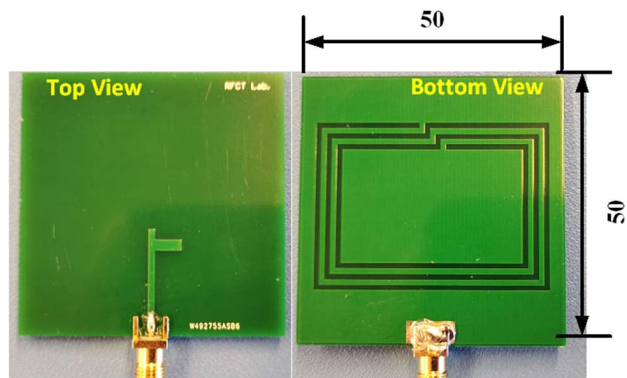

Fig. 9. Fabricated prototype of the proposed UCSMS on FR4 substrate ($\varepsilon_r = 4.3$, $tan\delta = 0.025$, $h = 1.6$ mm).

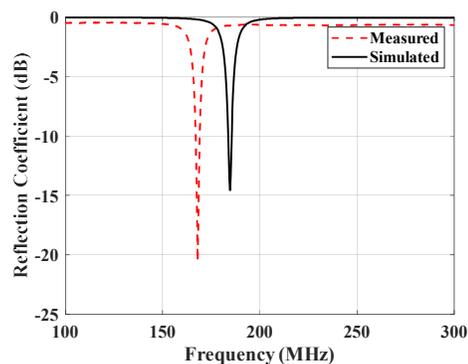

Fig. 10. Reflection coefficients of the unloaded UCSMS with 3-CSR.

To analyze the performance of the UCSMS, a measurement setup, as depicted in Fig. 11, was utilized. Pure washed fine sand from Bagged Product Supplies [49] was used to measure the frequency responses for different VWCs. Various types of soil exhibit different permittivity values at different VWCs, and the presence of various materials can also alter the permittivity of soil. For instance, the concentration of potassium chloride can increase the soil's permittivity. The proposed system requires calibration each time for different soil types and various material concentrations to establish the relationship between permittivity and VWC for each specific case. After calibration, the proposed UCSMS can measure VWC on a specific soil type with constant material concentration. Calibration of the system was conducted using pure sand without any impurities, as outlined in TABLE 3. The frequency responses of the proposed sensor were measured using a 4-port R&S VNA (VNA-ZVA40) and a standard SOLT (short, open, load, and through) calibration was performed before conducting each test. A cube of size $40 \times 40 \times 40$ mm³ is 3D printed and utilized as a soil container. The simulated and measured frequency responses of the proposed UCSMS are shown in Fig. 12. As the VWC in the soil increases, the frequency response of the UCSMS shifts leftwards with a significant frequency difference. The frequency of the sensor shifts from 158 MHz to 115 MHz as the VWC increases from 0 to 30 %. This frequency transition is used to measure the permittivity of the soil, which can be correlated to the VWC. The complex permittivity of the soil at 130 MHz for different VWCs is provided in TABLE 3. A close agreement between simulation and measurement results can be observed, validating the design of the proposed UCSMS.

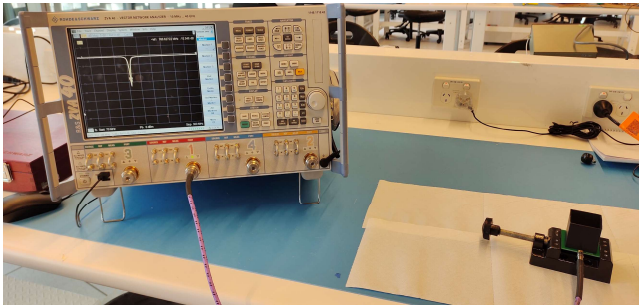

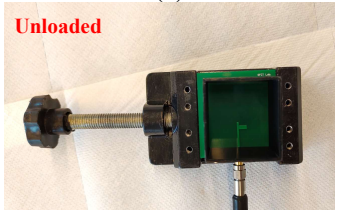

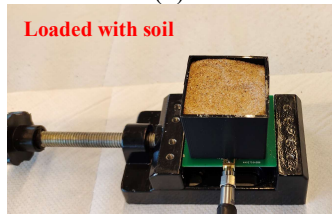

Fig. 11. Experimental setup to measure soil moisture, (a) measurement setup, (b) unloaded UCSMS, (c) loaded UCSMS with soil.

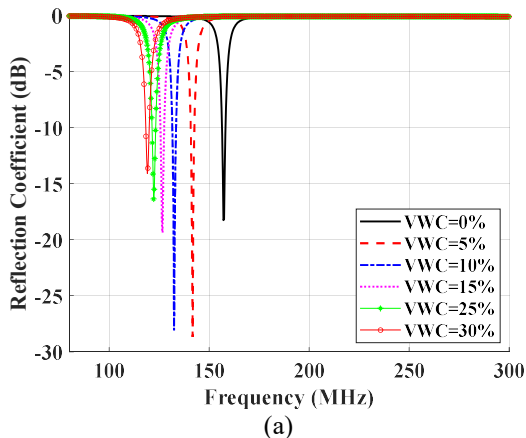

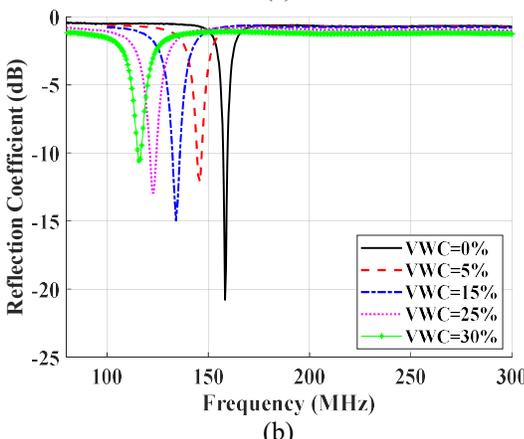

Fig. 12. Frequency responses of the proposed UCSMS with 3-CSR at different VWCs in the soil, (a) simulated, (b) measured.

TABLE 3. PERMITTIVITY OF SOIL FOR DIFFERENT VWCs AT 130 MHz [14]

| VWC (%) | 0 | 5 | 10 | 15 | 20 | 25 | 30 |
|---|---|---|---|---|---|---|---|
| $\varepsilon'_r$ | 2.5 | 6 | 8 | 14.5 | 18 | 21 | 23 |
| $\varepsilon''_r$ | 0.05 | 0.5 | 0.9 | 1.8 | 2.5 | 3.1 | 3.5 |

Three tests are performed separately to evaluate the precision of the measurement and frequency shifts have been measured with a maximum variation of ±5 MHz. The relationship between the frequency shift ($\Delta f$) and the complex permittivity is shown in Fig. 13, indicating a linear change.

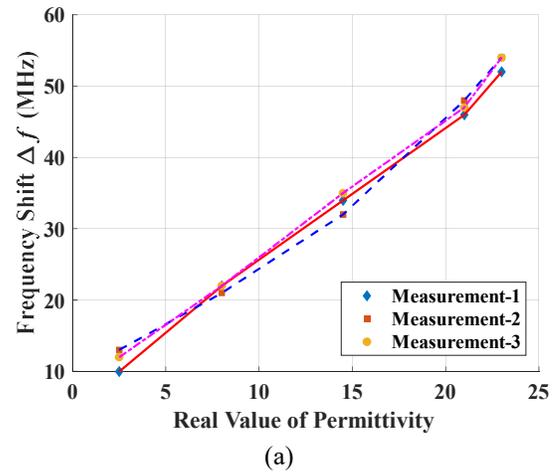

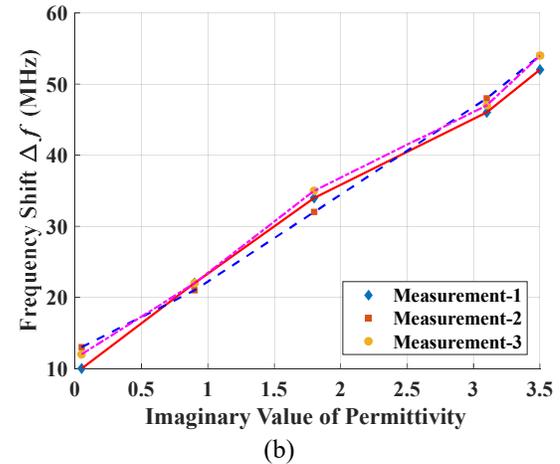

Fig. 13. Frequency shift vs permittivity, (a) real value, (b) imaginary value.

The proposed UCSMS is also analyzed with different heights of soil as depicted in Fig. 14(a) and corresponding frequency responses are measured. The frequency responses of the UCSMS for different heights are shown in Fig. 14(b), clearly indicating that the proposed UCSMS is not sensitive to the variation in the height of the soil under test. Hence, the proposed sensor can cover a larger VUT of the soil. To analyze the sensor's accuracy, two reference devices, Dielectric Assessment Kit (DAK 12) [50] and TDR-315L [51], were used to measure the permittivity and VWC of the soil. Fig. 15 displays the soil measurements using the DAK 12 and TDR-315L sensor.

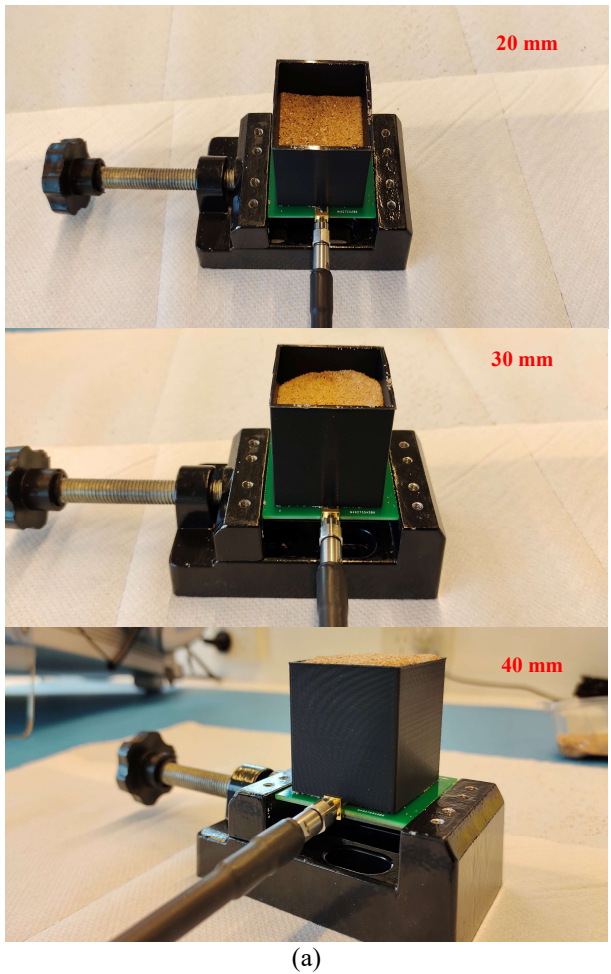

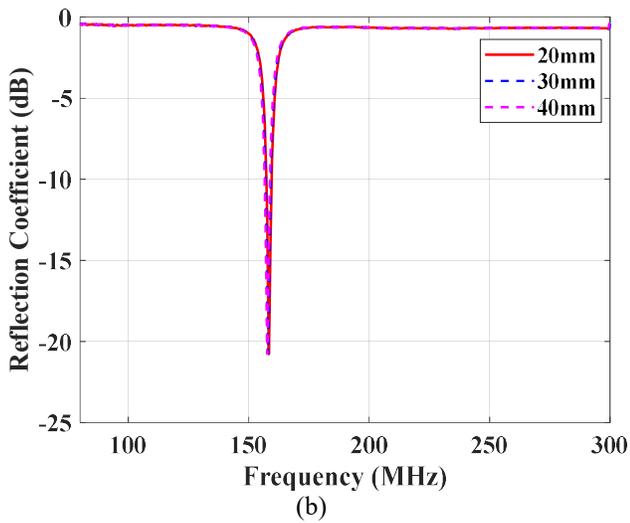

Fig. 14. Frequency Analysis for different soil heights, (a) different soil thicknesses, (b) frequency responses.

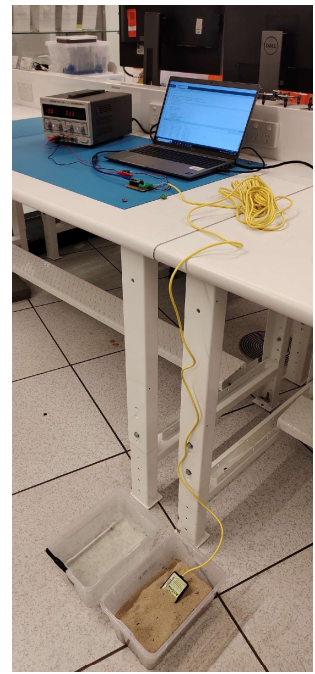

(a)

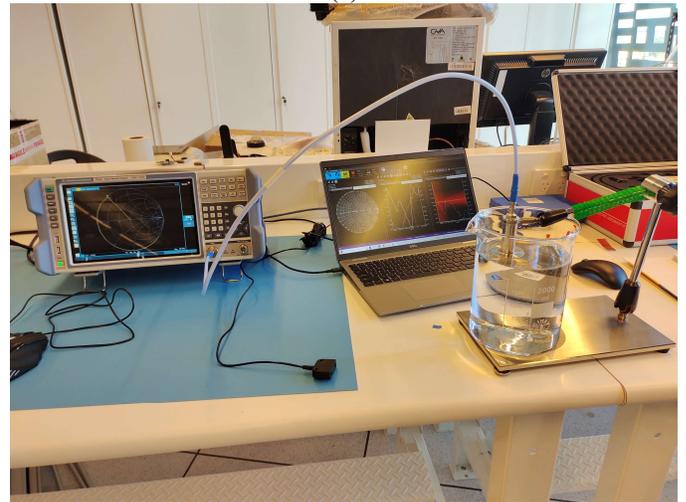

(b)

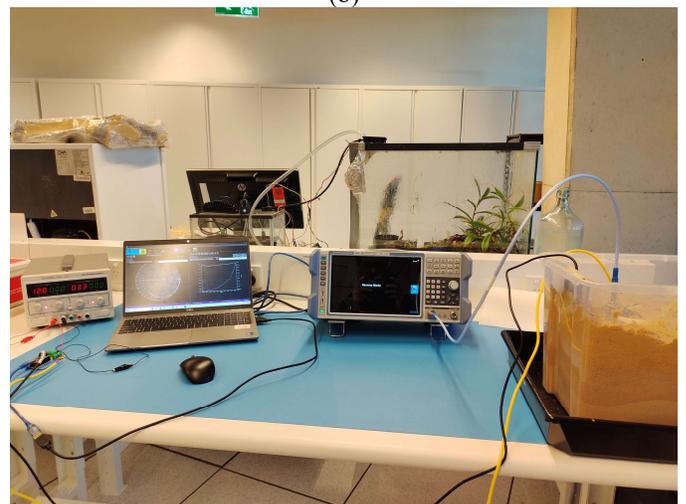

(c)

Fig. 15. Soil measurements using reference devices, (a) TDR-315L measurement, (b) DAK 12 calibration, (c) DAK 12 measurement.

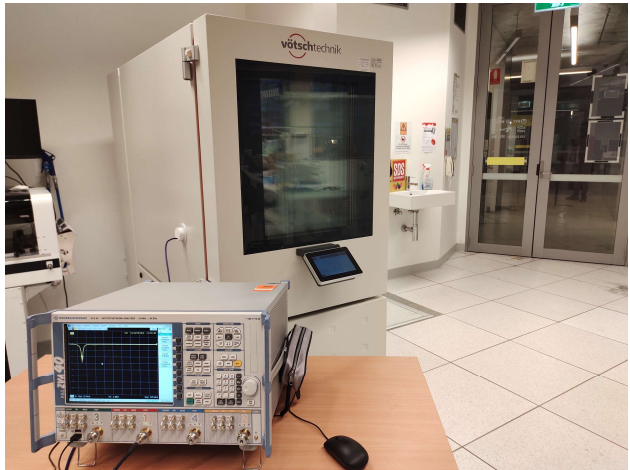

(a)

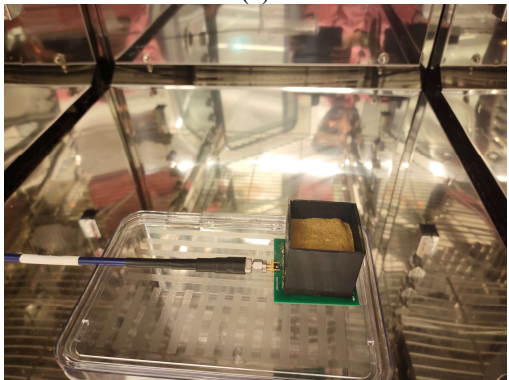

(b)

Fig. 16. Sensor performance analysis in the climate chamber, (a) test setup, (b) loaded UCSMS inside chamber.

To evaluate the sensor's performance in real-world conditions, additional tests have been conducted in a controlled vötschtechnik Environmental Testing Chamber (Temperature and Humidity) to measure the frequency responses under various environmental conditions. The laboratory setup for analyzing the sensor's performance under different environmental conditions is shown in Fig. 16. In a real-world environment, temperatures can drop as low as 0 °C during the nighttime and rise as high as 45 °C in summer, while humidity levels can vary from 40 % to 70 % at different times. To characterize the sensor in real-world conditions, temperature (0 to 45°C) and humidity (40 to 70%) are varied in the chamber to create extreme weather scenarios for 15 % and 30 % VWC values. For temperature variations, a constant humidity of 50 % is used, and for humidity measurements, a constant temperature of 25 °C is maintained. In the case of 30 % VWC, the measured results as depicted in Fig. 17, indicate that there is no significant difference in the frequency response readings with varying humidity values, as the sand is already saturated. However, for 15% VWC, the resonance frequency shifts towards the left side, indicating higher soil moisture with increased humidity. At 0 °C, where water exists in both liquid and solid forms, the VWC decreases due to the formation of ice, causing the resonance frequency to shift to the right. The frequency variation due to temperature is higher for 30 % VWC compared to 15% VWC. Similarly, at a high temperature of 45 °C, soil moisture decreases over time due to water vaporization, thereby validating the accuracy of the sensor. It should be noted that there is no significant difference in the resonance frequency between 10 and 30 °C.

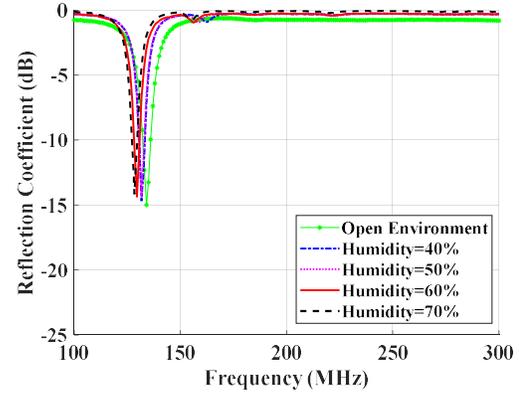

(a)

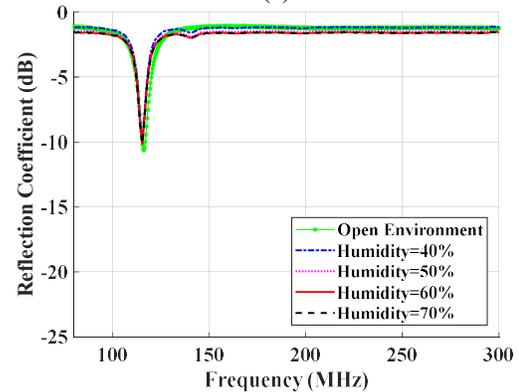

(b)

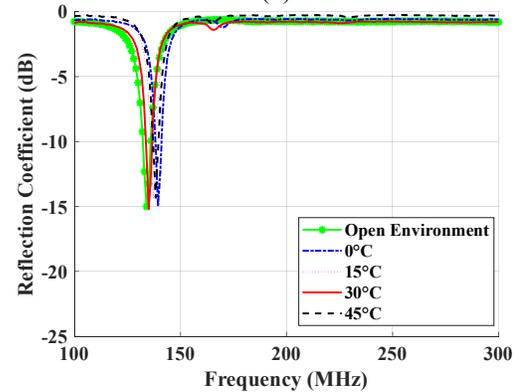

(c)

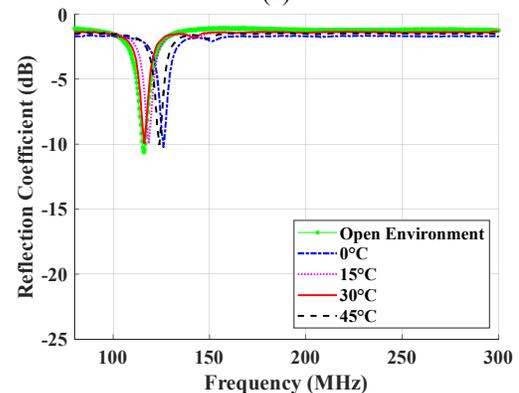

(d)

Fig. 17. Frequency response variations for different climate conditions in the environmental chamber, (a) 15 % VWC at 25 °C, (b) (a) 30 % VWC at 25 °C, (c) 15 % VWC at 50 % humidity, (d) 30 % VWC at 50 % humidity.

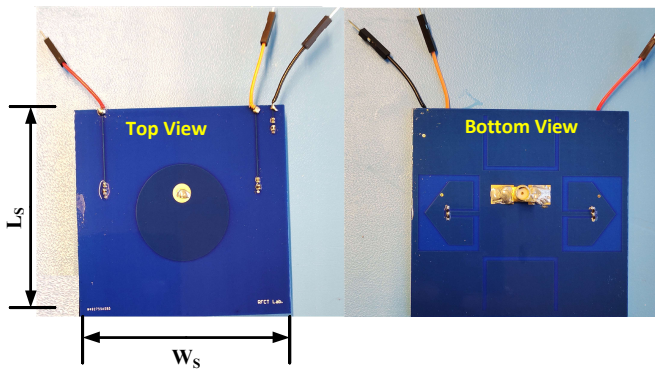

Fig. 18. Fabricated prototype of the proposed PRA on Rogers substrate ($\varepsilon_r = 3.55$, $\tan\delta = 0.0027$, $h = 1.524$ mm).

### B.  Simulated and Measured Results of PRA:

The PRA is designed and simulated using CST MWS 2019. The optimized structure is fabricated on a Rogers RO4003C substrate, and the fabricated prototype is illustrated in Fig. 18. The Skyworks SMV1231-079LF varactor diode is utilized, offering a capacitance range of $0.466 - 2.35$ pF. In the simulation, a non-linear diode model has been utilized and various capacitance values have been achieved using a parametric sweep. A biasing network is designed to bias the diodes which consists of two RF chokes, each with a value of 390 nH, and two resistors with a resistance of 1.2 kΩ, connected in series with the varactor diode.

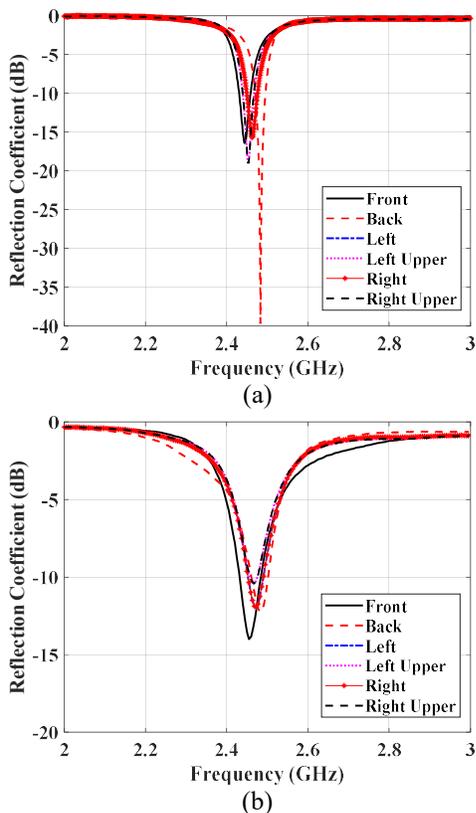

Fig. 19. Reflection coefficients of the proposed PRA, (a) simulated, (b) measured.

The simulated and measured reflection coefficients of the proposed PRA are shown in Fig. 19. The antenna resonates at 2.45 GHz, with a minimum measured bandwidth of 20 MHz for the 'upper left' and 'upper right' configurations. There is a slight variation in the resonance frequency of the PRA under different bias conditions of varactor diodes, with a maximum difference of 20 MHz between the 'front' and 'back' resonances. The simulation and measurement results exhibit good agreement, validating the successful realization of the antenna.

To analyze the far-field parameters, 2D radiation patterns of the proposed PRA are measured in an anechoic chamber at different basing voltages. A DC power supply was used to bias the varactor diodes to achieve various biasing conditions, and a standard gain horn antenna was utilized to measure the gain of the proposed PRA. The measurement setup for evaluating the far-field characteristics is shown in Fig. 20 and the simulated and measured 2D radiation patterns of the proposed PRA at 2.45 GHz are displayed in Fig. 21. Due to the presence of a DC power supply and biasing leads connected to the antenna for varactor diodes biasing, noise is introduced in the 'left' and 'right' radiation patterns. Nevertheless, for other biasing conditions, Fig. 21 illustrates a close agreement between the simulated and measured radiation patterns. Six different radiation patterns have been achieved with different biasing voltages across the diodes. Due to the symmetry of the structure in the *x-axis*, the patterns are symmetrical in the *xz-plane* as shown in Fig. 21. The patterns in the *yz-plane* are directional, with a constant $0^0$ lobe for all biasing conditions, as there are no varactor diodes across the U-shaped slots. The gains and directions of the maximum lobe in the *xz-plane* are summarized in TABLE 4. The antenna achieves a maximum measured gain of 5.63 dBi in the 'front' case. The directions of the maximum lobes are $6^0$, $185^0$, $10^0$, $25^0$, $340^0$, and $254^0$ for 'front', 'back', 'left', 'upper left', 'right', and 'upper right', respectively.

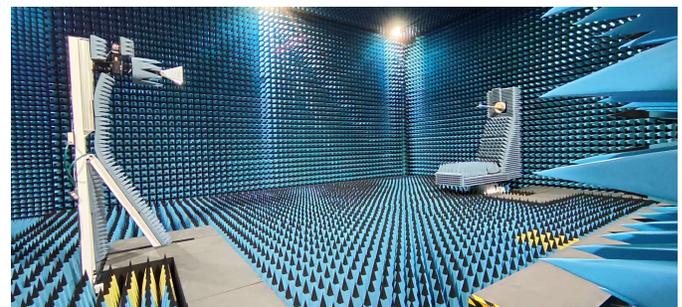

Fig. 20. Measurement setup to measure the far-field characteristics.

### C.  Figure of Merit

To compare the performance of the proposed sensor, the sensitivity of the UCSMS is calculated using (6),

$$Sensitivity = S = \left| \frac{f_1 - f_2}{f_u (\varepsilon_{r1} - \varepsilon_{r2})} \right| \times 100 \qquad (6)$$

where $f_u$ is the resonance frequency of the unloaded sensor, $f_1$ and $f_2$ are the resonance frequencies due to different materials, $\varepsilon_{r1}$ and $\varepsilon_{r2}$ represent relative permittivity of materials at $f_1$ and $f_2$, respectively. Additionally, a figure of merit (FOM) is defined in (7) taking into account the sensitivity ($S$), maximum electrical length of the sensor ($l$), and maximum measurable permittivity ($\varepsilon_{rm}$).

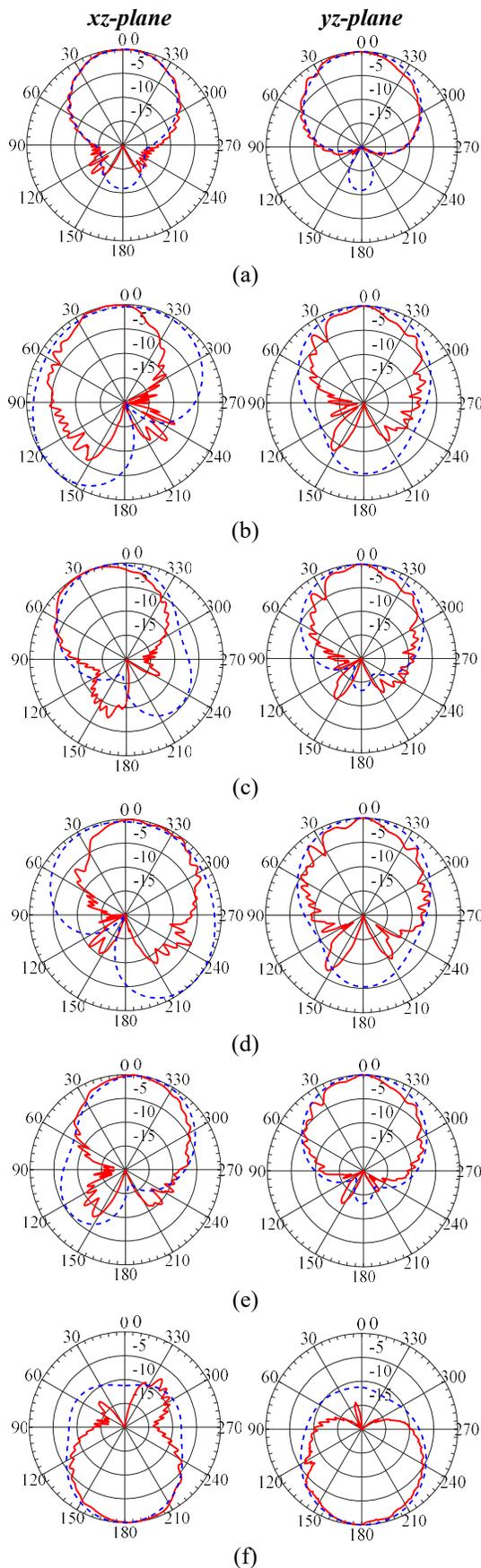

Fig. 21. Simulated and measured 2D radiation patterns at 2.45 GHz, (a) front, (b) left, (c) upper left, (d) right, (e) upper right, (f) back (solid line represents measured and the dotted line represents simulated plots).

TABLE 4. FAR-FIELD CHARACTERISTICS OF THE PROPOSED PRA

| Pattern Type | Gain (dBi) | | Maximum Lobe Direction ($\theta_m$) in the $xz$ plane | |
|---|---|---|---|---|
| | simulated | measured | simulated | measured |
| Front | 6.17 | 5.63 | $0^0$ | $6^0$ |
| Back | 4.45 | 4.14 | $180^0$ | $185^0$ |
| Left | 2.7 | 2.54 | $132^0$ | $10^0$ |
| Upper left | 5.117 | 5.08 | $15^0$ | $25^0$ |
| Right | 2.73 | 2.62 | $220^0$ | $340^0$ |
| Upper right | 5.08 | 4.923 | $345^0$ | $354^0$ |

$$Figure\ of\ Merit = \frac{S \times \varepsilon_{rm}}{l} \tag{7}$$

Based on these evaluations, a comparison of the proposed UCSMS with the reported sensors is summarized in TABLE 5. Furthermore, the proposed system possesses the ability to communicate with the base station to transfer information.

## IV. CONCLUSION

In this paper, we present a joint sensing and communication system for smart agriculture. The proposed system consists of an ultra-compact sensor for soil moisture measurement and a PRA for communication. An MCSR is used with a microstrip transmission line to achieve miniaturization. The proposed UCSMS operates at low frequencies, 180 MHz for 3-CSR, 102 MHz for 4-CSR, and 86 MHz for 5-CSR, making it suitable for covering a large volume of soil. The PRA operates at the 2.45 GHz WLAN band, facilitating the transmission of information to the base station. Integration of four varactor diodes with the communication antenna enables pattern reconfiguration, leading to the generation of six distinct radiation patterns with different bias conditions. This feature makes the system suitable for smart agriculture across diverse geographical landscapes. In standby mode, the PRA can also be utilized for WPT and EH applications to store power in a battery. This stored power can be utilized to bias the diodes to achieve reconfiguration. Both the UCSMS with 3-CSR and the PRA have been fabricated and measured, demonstrating a close agreement between the simulated and measured results. The sensor is adaptive and capable of measuring the permittivity of various materials within the range of 1–23.

## ACKNOWLEDGMENT

This project was supported by funding from Food Agility Cooperative Research Centre (CRC) Ltd, funded under the Commonwealth Government CRC Program. The CRC Program supports industry-led collaborations between industry, researchers and the community.

TABLE 5. COMPARISON OF THE UCSMS (ULTRA-COMPACT SOIL MOISTURE SENSOR) WITH REPORTED SENSORS

| Ref. | Size at $f_a$ ($\lambda_0{}^2$) | $f_a$ (GHz) | Number of Sensing Bands | Measurement Technique | Sensitivity @ $max(\varepsilon_r)$ (%) | Max Measured Permittivity | FOM $\frac{1}{\lambda_0}$ |
|---|---|---|---|---|---|---|---|
| This Work | 0.028×0.028 | 0.170 | 1 | MCSR (3-CSR) | 2.05 | 23 | 1683.92 |
| [4] | 0.67×0.13 | 4 | 2 | SRR and CSRR | 0.9 | 16.7 | 22.43 |
| [23] | 0.35×– | 2.67 | 3 | CSRR | 1.6 | 9.2 | 42.05 |
| [24] | – | 2.4 | 1 | CSRR | 0.19 | 79.5 | - |
| [25] | 0.32×0.2 | 2.45 | 1 | M-CSRR | 0.2 | 70 | 43.75 |
| [26] | 0.36×– | 2.7 | 1 | CSRR | 1.7 | 10.2 | 48.167 |
| [29] | 0.34×0.034 | 1.017 | 1 | Shorted-Dipole | 0.614 | 19 | 34.31 |
| [31] | 0.42×0.44 | 4.7 | 1 | Frequency Selective Multipath Filter | 0.214 | 26 | 12.65 |
| [30] | 0.198×0.198 | 2.38 | 1 | EBG Resonator | 0.224 | 70 | 79.19 |
| [27] | 0.184×0.372 | 2.234 | 1 | SRR | 0.04476 | 70 | 8.42 |
| [28] | 0.6×0.4 | 3.49 | 1 | Complementary Curved Ring Resonator | 4.47 | 4.4 | 32.78 |
| [5] | 0.38×0.38 | 0.56 | 1 | Metamaterial Absorber | 0.109 | 19.1 | 5.47 |